\documentclass[letterpaper, 11pt]{article}
\pdfoutput=1
\usepackage{amsbsy}
\usepackage{amsfonts}
\usepackage{amsmath}
\usepackage{amssymb}
\usepackage{amsthm}
\usepackage[page]{appendix}
\usepackage{bm}
\usepackage{color}
\usepackage{epsfig}
\usepackage{geometry}
\usepackage{graphicx}
\usepackage{hyperref}
\usepackage{subfigure}
\usepackage{xspace}

\usepackage[super, comma, sort&compress]{natbib}

\usepackage[version=3]{mhchem}
\usepackage{amssymb}
\usepackage{amsmath}
\usepackage{epstopdf}
\usepackage{subfigure}

\usepackage[table,xcdraw]{xcolor}
\usepackage{xr}
\usepackage{pgfplots}
\usepackage{tikz}

\usetikzlibrary{calc, shapes, arrows, positioning, angles, quotes,
	patterns, fit}

\pgfplotsset{compat=newest} 
\pgfplotsset{plot coordinates/math parser=false} 
\newlength\figureheight 
\newlength\figurewidth

\title{Diagnostics of Data-Driven Models: Uncertainty Quantification of PM7 Semi-Empirical Quantum Chemical Method}

\author{James Oreluk\footnote{Department of Mechanical Engineering, University of California at Berkeley, Berkeley, California 94720-1740}, Zhenyuan Liu\footnotemark[1], Arun Hegde\footnotemark[1], Wenyu Li\footnotemark[1], \\Andrew Packard\footnotemark[1], Michael Frenklach\footnote{Corresponding author: \href{mailto:frenklach@berkeley.edu}{frenklach@berkeley.edu}, Department of Mechanical Engineering, University of California at Berkeley, Berkeley, California 94720-1740}, Dmitry Zubarev\footnote{IBM Almaden Research Center, 650 Harry Road, San Jose, California 95136} }

\begin{document}
	
	\maketitle
	
\begin{abstract}
	We report an evaluation of a semi-empirical quantum chemical method PM7 from the perspective of uncertainty quantification. Specifically, we apply Bound-to-Bound Data Collaboration, an uncertainty quantification framework, to characterize a) variability of PM7 model parameter values consistent with the uncertainty in the training data, and b) uncertainty propagation from the training data to the model predictions. Experimental heats of formation of a homologous series of linear alkanes are used as the property of interest. The training data are \textit{chemically accurate}, i.e., they have very low uncertainty by the standards of computational chemistry.  The analysis does not find evidence of PM7 consistency with the entire data set considered as no single set of parameter values is found that captures the experimental uncertainties of all training data. Nevertheless, PM7 is found to be consistent for subsets of the training data. In such cases, uncertainty propagation from the chemically accurate training data to the predicted values preserves error within bounds of chemical accuracy if predictions are made for the molecules of comparable size. Otherwise, the error grows linearly with the relative size of the molecules.
\end{abstract}

\section{Introduction}
Uncertainty quantification (UQ) is a critical diagnostic for the models that are positioned as a source of actionable predictions. Computational models of this kind are abundant in chemistry. One can think of kinetic models,\cite{doi:10.1146/annurev.pc.41.100190.002021, C0CS00207K, frenklach2007transforming} classical force-fields, \cite{doi:10.1021/acs.chemrev.5b00505, doi:10.1021/ct700127w} semi-empirical Hamiltonians,\cite{doi:10.1021/acs.chemrev.5b00584, doi:10.1021/jp402719k, doi:10.1002/wcms.1161} approximate density functionals \cite{doi:10.1063/1.4704546, doi:10.1021/ct6001016, doi:10.1021/jp801805p, doi:10.1021/ct0502763, C4CP00908H} and the blossoming field of machine learning/deep learning models.\cite{doi:10.1016/j.febslet.2015.10.003, doi:10.1021/ci400187y, schutt2017quantum, hegde2017machine, doi:10.1002/jcc.24764, doi:10.1063/1.4986081}

One of the challenges of the model selection and curation is their tendency to proliferate uncontrollably if there is flexibility in the model's form and fitting to reference data is involved. The rich landscape of approximate density functionals \cite{medvedev2017density} is a case to the point. Models produced within data-centric framework that are application-oriented will inevitably face the same challenge at a much larger scale.\cite{doi:10.1021/jm4004285, doi:10.1080/1062936X.2016.1253611}

Currently, models that require parameter fitting are compared on the basis of some form of the fitting error.\cite{peverati2014quest} Our goal here is to develop a framework for model-data diagnostics, that informs users about the model from the point of view of the UQ. First, given a set of training data with known, presumably low, uncertainty, we want to be able to establish if there is a single contiguous domain of the model parameters that captures all training data within their uncertainty.  Second, knowing the domain of parameter values that are consistent with the training data, we want to establish how the variance in the parameter values propagates into the uncertainty of the model predictions. 

There is a noticeable growth of interest in the assessment of uncertainty of the models used in computational chemistry (see e.g., Refs. \cite{russi2010uncertainty, faver2012effects, yang2018atomic, simm2018error} and citations therein). A recent analysis of the error assessment in computational chemistry arrives at the following conclusion:\cite{Simm:2017:0009-4293:202} ``$\ldots$ a procedure for quantifying the uncertainty associated with computational models, in particular with quantum chemical calculations, is mandatory despite their first-principles character. Otherwise, it may be difficult to draw meaningful conclusions. Unfortunately, this procedure is neither well established nor straightforward.'' Earlier we demonstrated how applying the Bound-to-Bound Data Collaboration framework \cite{frenklach2002prediction, seiler2006numerical, russi2010uncertainty} for UQ can be incorporated into DFT framework in order to quantify and improve uncertainty of the predictions.\cite{PhysRevLett.112.253003}  In the present study, we apply the same framework as a diagnostic tool for semi-empirical quantum chemical model PM7.\cite{stewart2013optimization}


Semi-empirical model chemistries are popular because of the low computational overhead that makes it possible to tackle large systems in a relatively short time. Interpretable mechanistic nature of these methods and straight-forward manner in which empirical data can be incorporated are the reasons why semi-empirical quantum chemistry finds a wide range of ubiquitous applications and sees ongoing development.\cite{dral2016semiempirical1, dral2016semiempirical} 

The goal of the presented study is to determine if a) there is a single set, i.e., feasible set, of PM7 parameter values that satisfies uncertainty bounds of the training data, and b) it is possible to achieve predictions outside the training data with the uncertainty as low as the uncertainty of the training data.

\section{Methods and Data}

\subsection{Bound-to-Bound Data Collaboration}

This section provides a brief overview of Bound-to-Bound Data Collaboration (B2BDC), with a focus on the methodological details specific to the present study. A more detailed discussion can be found elsewhere.\cite{frenklach2002prediction, seiler2006numerical, russi2010uncertainty, feeley2004consistency, frenklach2016comparison} 

B2BDC is a deterministic framework for uncertainty quantification, where uncertain parameters are constrained by combining models and experimental data. Let $\{M_e(x)\}_{e=1,\ldots,N}$ denote a collection of models mapping from a common parameter space to various scalar-valued quantities of interest (QOIs). For a given parameter vector $x \in \mathbb{R}^n$, the expression $M_e(x)$ evaluates the model prediction for $e$-th QOI. Experimental observations of the QOIs come in B2BDC in the form $\{[L_e,U_e ]\}_{e=1,\ldots,N}$, where the uncertainty is characterized by bounds. Oftentimes, there is prior knowledge to confine the parameters, $x$, to a set $\mathcal{H} \subset \mathbb{R}^n$. Constrained by both prior knowledge on $x$ and experimental uncertainty ranges, each QOI has an associated feasible set of parameters for which model evaluations agree with the corresponding experimental bounds:
\begin{equation} \label{eq:feasibleE}
\mathcal{F}_e = \{x \in \mathcal{H}: L_e \leq M_e(x) \leq U_e  \}, \;\; e = 1,2,\ldots, N.
\end{equation}
Naturally, one is often interested in the set of parameters which satisfies multiple observations: 
\begin{equation} \label{eq:feasibleI}
\mathcal{F}_I = \cap_{e\in I}\mathcal{F}_e,
\end{equation}
where $I \subset \{1,2, \ldots, N \}$ is the index set of the QOIs. For example, $\mathcal{F}_{1:3,5}$ denotes the feasible set associated with QOIs, $e = 1,2,3,5$. If the set $\mathcal{F}_{1:N}$ is nonempty, the models and data are said to be \textit{consistent}. If the models can be accurately represented by polynomial surrogates, this notion can be efficiently and provably quantified through constrained optimization.\cite{feeley2004consistency} A polynomial surrogate is a model which mimics the behavior of an underlying simulation model as close as possible, while being computationally cheaper to evaluate.\cite{box1987empirical} In cases where the models are not well characterized by simple surrogates but are cheap to evaluate, direct sampling can provide a means of assessing consistency. To prove consistency, we must find at least one parameter vector $x$ in $\mathcal{F}_{1:N}$. However, this becomes very challenging when the parameter space is large and there are many QOI models whose experimental bounds must be satisfied.

In B2BDC, the prediction for a particular model, say $M_p(x)$, amounts to establishing bounds on the range of the model prediction, subject to model parameters being within a feasible set, $\mathcal{F}_I$:
\begin{equation} \label{eq:prediction}
\left[\min_{x \in \mathcal{F}_I} M_p(x), \; \max_{x \in \mathcal{F}_I} M_p(x) \right].
\end{equation}
The prediction model need not be a member of the collection of models $\{M_e(x)\}_{e=1,\ldots,N}$. In certain cases, Eq. \ref{eq:prediction} can be efficiently solved via constrained optimization.\cite{seiler2006numerical} Note, inner approximations to this interval can be found by taking samples of $\mathcal{F}_I$ and evaluating the prediction model for the collected sample set. The inner approximation can be interpreted as saying the prediction model can be no more predictive than the computed bounds and wider intervals indicate greater uncertainty in the prediction. 

\subsection{Selection of QOIs and experimental data for PM7}

There are a multitude of approaches available for validating a computational model to experimental data. The two most important factors when validating a model is the choice of validation data and the specific experimental feature to validate the model by.

In this study, we will be considering a single homologous series, namely linear alkanes. We chose this homologous series for its simplicity, accumulated knowledge, and the fact that the number of uncertain model parameters does not increase with the size of the molecule. There are 27 adjustable model parameters in PM7 which are used to parameterize the interactions of carbon and hydrogen atoms. Given the problem choice, we should expect the ``best case scenario'' from the results.

For a given parameterization, PM7 as a model provides many output responses, e.g., vibrational frequencies, ionization potential, homo-lumo energies, heat of formation, etc. Due to the sheer number of model responses, the availability of accurate experimental data will be used to guide the choice of a quantity of interest. We also keep in mind that the primary output value of PM7 is standard heat of formation. Ruscic et al, \cite{ruscic2017active} employing a thermochemical network approach, reported accurate thermochemical values for a set of molecules, including members of linear alkanes, methane to octane (Table \ref{table:expData}). Uncertainty listed in Table \ref{table:expData} represents the reported 95\% confidence interval. In present study, the heats of formation of linear alkanes will be used as our QOIs. The bounds used are the reported heat of formation $\pm$ the uncertainty. Table \ref{table:expData} also shows the QOI models used in the study, namely the heat of formation of methane, ethane to octane. For sake of simplicity, each QOI model is indexed by the number of carbon atoms.

\begin{table}[!htb]
	\centering
	\caption{List of the QOI models $M_e$, the associated index $e$ (equivalent to the number of carbon atoms in the alkane), the reported heat of formation, and uncertainty \cite{ruscic2017active}}
	\label{table:expData}
	\begin{tabular}{cccc}
		\textbf{$e$} & \textbf{$M_e$} & \textbf{$\Delta_\text{f}$H$^\circ_{298}$ (kcal/mol)} & \textbf{Uncertainty} \\ \hline
		\rowcolor[HTML]{FFFFFF} 
		1 & $\Delta_\text{f}$H$^\circ_{298}(\ce{CH4})$ & -17.8107 & $\pm 0.0136$ \\
		2 & $\Delta_\text{f}$H$^\circ_{298}(\ce{C2H6})$ & -20.0550 & $\pm 0.0335$ \\
		\rowcolor[HTML]{FFFFFF} 
		3 & $\Delta_\text{f}$H$^\circ_{298}(\ce{C3H8})$ & -25.1028 & $\pm 0.0454$ \\
		4 & $\Delta_\text{f}$H$^\circ_{298}(\ce{C4H10})$ & -30.1004 & $\pm 0.0645$ \\
		\rowcolor[HTML]{FFFFFF} 
		5 & $\Delta_\text{f}$H$^\circ_{298}(\ce{C5H12})$ & -34.9761 & $\pm 0.0741$ \\
		6 & $\Delta_\text{f}$H$^\circ_{298}(\ce{C6H14})$ & -39.8972 & $\pm 0.0813$ \\
		\rowcolor[HTML]{FFFFFF} 
		7 & $\Delta_\text{f}$H$^\circ_{298}(\ce{C7H16})$ & -44.8207 & $\pm 0.1147$ \\
		8 & $\Delta_\text{f}$H$^\circ_{298}(\ce{C8H18})$ & -49.7777 & $\pm 0.1601$
	\end{tabular}
\end{table}

\subsection{Motivation for direct sampling}

B2BDC typically employs surrogate models to represent an underlying simulation. The use of surrogates makes tasks like optimization and sensitivity analysis much more efficient due to inexpensive evaluations. When the surrogates are polynomials, consistency and prediction can be addressed efficiently and provably. However, we have found constructing surrogates for PM7 particularly challenging for two reasons, as we will detail below with the $\Delta_{\text{f}}$H$^\circ_{298}(\ce{C4H10})$ serving as our example. 

First, we do not have a well-defined prior region $\mathcal{H}$ to confine our search for feasible parameters as no bounds are specified for these parameters. To address this, we constructed an initial $\mathcal{H}$ by centering the search around the PM7 nominal parameter vector, \cite{stewart2013optimization} $x_{\text{nom}}$, which is listed in Table \ref{table:nominalparameter} of the Supporting Information. An initial search region was constructed by selecting perturbations $\pm \delta_i$ to the $i$-th parameter with all other parameters remained fixed such that the interval defined by $x_{\text{nom},i} + \delta_i$ produced a 10 kcal/mol change in $\Delta_{\text{f}}$H$^\circ_{298}$. We denote this initial region by $\mathcal{H}_1 = [ x_{\text{nom}} \pm \delta ]$, as shown in Table \ref{table:nominalparameter}. 

\begin{figure}[!htb]
	\centering
	\includegraphics[width=0.6\textwidth]{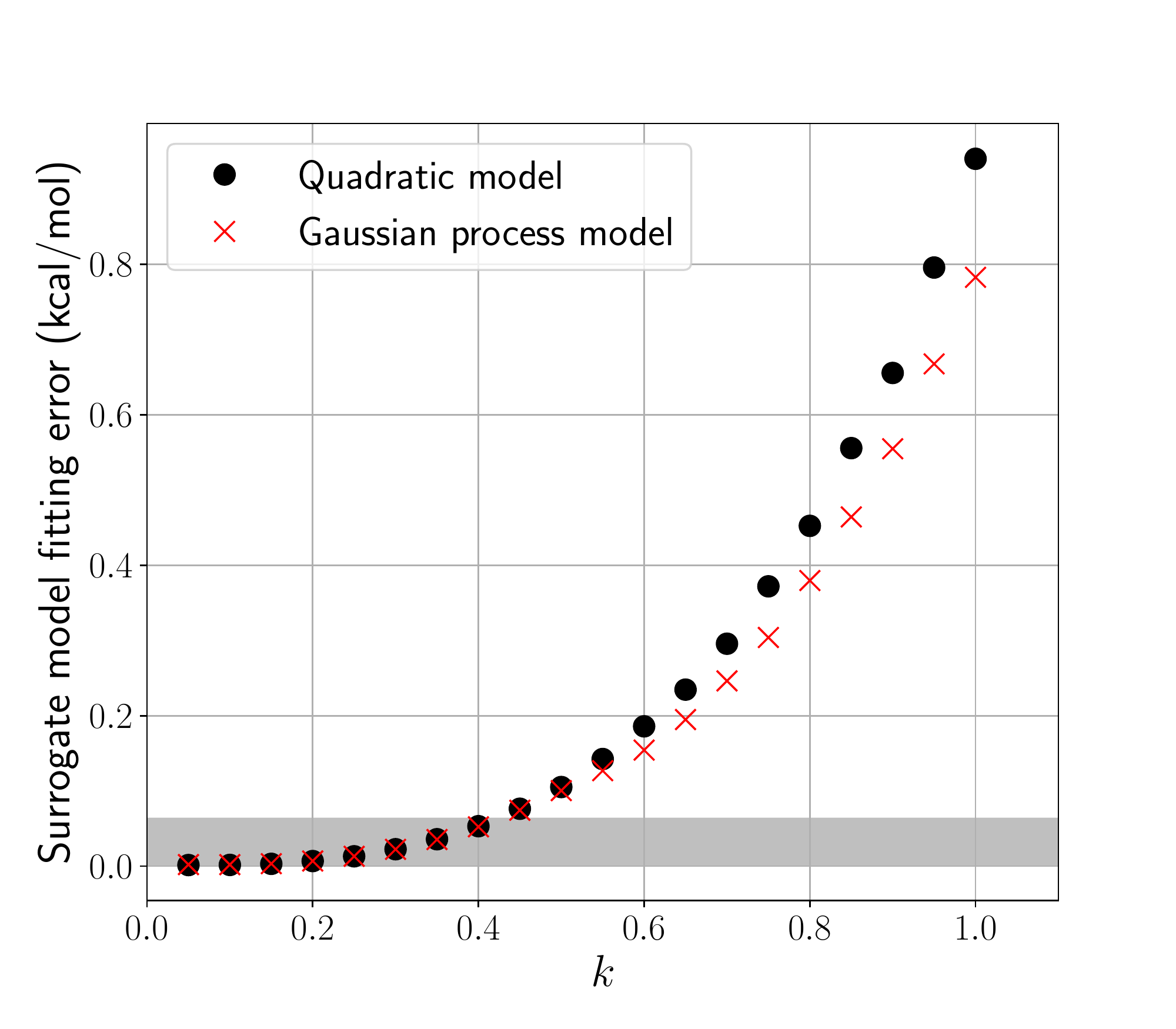}%
	\caption{Surrogate model fitting error for the heat of formation of \ce{C4H10} evaluated on a shrinking domain $\mathcal{H}_k = [ x_{\text{nom}} \pm k \times \delta]$, with $k \in [0, 1]$. The nominal PM7 value, $x_{\text{nom}}$, and $\delta$ are reported in Table \ref{table:nominalparameter}. As the domain shrinks, by reducing $k$, the surrogate model fitting error decreases to a point where the error falls below the experimental uncertainty (shown in grey) of 0.0645 kcal/mol. Comparison between a quadratic surrogate (black dots) and a Gaussian process model (red crosses) with a constant mean and a squared exponential covariance function is shown. 10-fold cross-validation was used to estimate the fitting error. }
	\label{fig:surrogateModelErrors}%
\end{figure}

Second, our experience has shown that surrogate models are only accurate (to within the experimental uncertainty) on small domains. To illustrate this, we compare the fitting error of a quadratic polynomial and Gaussian process surrogate on a shrinking domain $\mathcal{H}_k = [ x_{\text{nom}} \pm k \times \delta ]$ where $k \in [0, 1]$, as shown in Figure \ref{fig:surrogateModelErrors}. For each $k$, 7500 Latin Hypercube samples were generated in $\mathcal{H}_k$ to construct the surrogates. The Gaussian process was implemented by using MATLAB's \textit{fitrgp} function.\cite{matlab} In order to attain fitting error below our target, the experimental uncertainty, the initial region needed to be reduced by the factor of $k = 0.4$. The reduced domain $\mathcal{H}_{0.4}$ is $5.5 \times 10^{10}$ times smaller by volume as compared to the original domain $\mathcal{H}_1$. Based on this result, we concluded that it would require an intractable number of surrogate models to explain the behavior of PM7 over the entire $\mathcal{H}$ within the desired accuracy. 

\subsection{Direct Sampling for Consistency}

In light of the challenges discussed in the preceding section, we approached consistency and prediction by directly sampling $\mathcal{H}$. Still, despite PM7's fast evaluation time, particularly for fixed geometries, evaluating consistency for multiple QOIs remains a computational challenge. We break this challenge into two tasks: (1) allocating computational resources to sample $\mathcal{H}$; and (2) determining a prior region $\mathcal{H}$ to confine our search for feasible parameters.

\subsubsection{Computational burden of direct sampling PM7}

One of the limitations encountered in the present study was the number of samples we could evaluate for PM7 with the computational resources available. Evaluating PM7 for millions of parameter combinations is an embarrassingly parallel task, but computational efficiency can be improved by reducing an I/O bottleneck occurring in the evaluation process.  

Throughout the study, the input molecular geometries were obtained via optimization with the PM7 nominal parameter vector, which remained frozen afterwards, during the subsequent direct sampling. The freezing of the molecules geometries was done to avoid any unphysical reorganization or distortion. 

The computational chemistry program MOPAC \cite{stewart2012colorado} was used for evaluation of PM7 parameter sets, and the following steps were taken. First, an external file was created which assigned new values to the 27 uncertain model parameters. Second, an input file was created that specified the geometry of the molecule, run-time options, and the location of the external file. Once both the external file and input file are created, MOPAC was executed and PM7 was evaluated. An output text file of the results was generated by MOPAC at the end of the execution. The output file was parsed and the quantity of interest, i.e., the heat of formation was extracted. This process was repeated for each parameter vector and for each alkane.

One of the practical aspects to be solved in working with MOPAC is handling efficiently multiple reading and writing of text files for each execution, which creates a significant bottleneck in I/O when parallelizing the direct sampling approach. To reduce I/O time, a memory based file system was used, that enabled all files to be read and written into system memory instead of a traditional storage device. For example, using a Seagate 7200 RPM SATA drive, read and write speeds were 204.0 and 194.9 MB/s, respectively. Using a memory based file system we could read and write at 6020.7 and 5840.6 MB/s, respectively. To relate this to sampling, the time to evaluate $10^4$ different parameter values for \ce{C4H10} on a workstation equipped with an Intel Core i7-6700K processor took 305.0 seconds using the Seagate hard drive and 215.7 using the memory based file system, a speed up of 41\%. This speedup in I/O directly translates into more evaluations of PM7 per second.  

The MATLAB Parallel Computing Toolbox \cite{matlab} was used to write input files, execute PM7, and read output files in parallel. Two workstations, each with an Intel Core-i7 6700K processor were used to directly sample PM7 and evaluate the heats of formation of a series of alkanes, methane to nonane.  

\subsubsection{Determination of a prior search region}
The choice of the prior parameter domain, $\mathcal{H}$, determines all subsequent analysis and hence it needs to be well motivated. Our goal is to assess the predictivity of PM7 for large alkanes while being consistent with experimental data available for small alkanes. Due to limited data, we set aside heptane and octane, the two largest alkanes for which we have experimental bounds, to assess predictivity. Conversely, we also set aside methane and ethane so as not to bias the result. Although the physical properties of methane and ethane may be representative of the homologous series, the chemical properties, i.e., the energetics, are very different as seen through group additivity.\cite{benson1969additivity} Propane and butane were also not included in the initial region search, since a priori it was not known if a feasible set exists. Based on these considerations, we prioritized our determination of $\mathcal{H}$ on feasible samples for pentane and hexane, $\mathcal{F}_{5:6}$.

A non-linear programming solver, \textit{fmincon}, \cite{matlab} was used to minimize the following objective function, $\left\| M_{5:6}(x) - y_{5:6} \right\|_2^2$, where, $M_{5:6}(x)$ are the heats of formation of pentane and hexane computed by PM7 at the parameter vector $x$. $y_{5:6}$ are the reported heats of formation for pentane and hexane from Table \ref{table:expData}. The local optimal parameter vector found, $x_\text{opt}$, was feasible for pentane and hexane, therefore $x_\text{opt} \in \mathcal{F}_{5:6}$. A volume was taken around $x_{\text{opt}}$ in order to collect samples of $\mathcal{F}_{5:6}$. For each sampled point, the heats of formation of pentane and hexane were computed by PM7. $2 \times 10^5$ samples were generated by employing a Latin Hypercube design from the volume $[x_{\text{opt}} \pm x_{\text{opt}} \times 1 \times 10^{-3}]$ and 65 samples were found to be within $\mathcal{F}_{5:6}$.

Principal component analysis (PCA) was conducted to identify a rotated coordinate system around the samples of $\mathcal{F}_{5:6}$, where all principal components were preserved. Considering the arbitrary choice of the volume used in the sampling of $\mathcal{F}_{5:6}$, each PCA direction was extended ten times of that of the $\mathcal{F}_{5:6}$ samples. This allowed for an even larger sample region to be considered. A Latin Hypercube design was used to generate uniform samples in the rotated and extended volume. 

In total, 5.76 million samples were generated uniformly within the PCA-rotated volume and were used to evaluate PM7 for the heats of formation of nine alkanes, \ce{CH4} to \ce{C9H20}. Shown in Table \ref{table:parameterRange} are the extreme parameter values from the generated samples. There were 164,569 samples which were feasible for at least one alkane in the training set. The number of feasible samples found quickly dropped when considering feasibility with multiple alkanes. For example, there were 19167 samples that were feasible with at least two alkanes, 6193 samples feasible with at least three alkanes, 3110 samples feasible with at least four alkanes, 1989 samples feasible with at least five alkanes, and 169 samples feasible with at least six alkanes.

\begin{table}[!htb]
	\centering
	\caption{Extreme parameter values from the search region.}
	\label{table:parameterRange}
	\begin{tabular}{lll}
		\textbf{Parameter} & \textbf{min(x)} & \textbf{max(x)} \\ \hline
		\rowcolor[HTML]{FFFFFF} 
		\textbf{USS$_\text{H}$} & -14.815 & -9.5475 \\
		\textbf{BETAS$_\text{H}$} & -10.444 & -6.383 \\
		\rowcolor[HTML]{FFFFFF} 
		\textbf{ZS$_\text{H}$} & 0.85244 & 1.5053 \\
		\textbf{GSS$_\text{H}$} & 10.953 & 15.822 \\
		\rowcolor[HTML]{FFFFFF} 
		\textbf{FN11$_\text{H}$} & 0.14962 & 0.23173 \\
		\textbf{FN21$_\text{H}$} & 1.2161 & 1.4849 \\
		\rowcolor[HTML]{FFFFFF} 
		\textbf{FN31$_\text{H}$} & 0.83875 & 1.0458 \\
		\textbf{ALPB$_\text{H}$} & 3.9629 & 4.8467 \\
		\rowcolor[HTML]{FFFFFF} 
		\textbf{XFAC$_\text{H}$} & 2.3077 & 2.7852 \\
		\textbf{USS$_\text{C}$} & -52.553 & -45.61 \\
		\rowcolor[HTML]{FFFFFF} 
		\textbf{UPP$_\text{C}$} & -43.679 & -37.301 \\
		\textbf{BETAS$_\text{C}$} & -15.285 & -11.804 \\
		\rowcolor[HTML]{FFFFFF} 
		\textbf{BETAP$_\text{C}$} & -9.2246 & -6.7715 \\
		\textbf{ZS$_\text{C}$} & 1.5914 & 2.2229 \\
		\rowcolor[HTML]{FFFFFF} 
		\textbf{ZP$_\text{C}$} & 1.501 & 2.0466 \\
		\textbf{GSS$_\text{C}$} & 10.314 & 13.662 \\
		\rowcolor[HTML]{FFFFFF} 
		\textbf{GSP$_\text{C}$} & 9.7585 & 13.092 \\
		\textbf{GPP$_\text{C}$} & 9.3945 & 12.464 \\
		\rowcolor[HTML]{FFFFFF} 
		\textbf{GP2$_\text{C}$} & 8.6452 & 11.014 \\
		\textbf{HSP$_\text{C}$} & 0.66956 & 0.77601 \\
		\rowcolor[HTML]{FFFFFF} 
		\textbf{FN11$_\text{C}$} & 0.045528 & 0.055036 \\
		\textbf{FN21$_\text{C}$} & 4.3632 & 5.1283 \\
		\rowcolor[HTML]{FFFFFF} 
		\textbf{FN31$_\text{C}$} & 1.4298 & 1.7154 \\
		\textbf{ALPB$_\text{H}$$_\text{C}$} & 0.76974 & 1.1501 \\
		\rowcolor[HTML]{FFFFFF} 
		\textbf{XFAC$_\text{H}$$_\text{C}$} & 0.15181 & 0.21488 \\
		\textbf{ALPC$_\text{C}$} & 2.3296 & 3.0615 \\
		\rowcolor[HTML]{FFFFFF} 
		\textbf{XFAC$_\text{C}$} & 0.71959 & 0.96953
	\end{tabular}
\end{table}

\section{Results}

\subsection{Feasible Set Search}
\label{section:fss}
For each of the eight alkanes, a feasible parameter vector could be found that would predict its heat of formation within the experimental bounds. Feasible parameters were also found for all pairwise combinations of alkanes, i.e., a parameter vector could be found that would predict the heat of formation for any two alkanes within their respective experimental bounds. The percentage of feasible samples found for each pair of alkanes in the training data set is reported in Table \ref{table:nFeasible} of the Supporting Information. Methane had the smallest percentage of feasible samples for a single alkane and the largest was for octane. This difference in percentages of feasible samples could be due to the uncertainty bounds associated to octane being nearly an order of magnitude larger than that of methane and the difference in molecular structure of methane compared to all others in the homologous series.

From the generated samples, there were 4020 samples of $\mathcal{F}_{5:6}$, the alkanes which we had focused our search on. However, from the generated samples, there was no single parameter vector that was able to simultaneously satisfy the uncertainty bounds of all the training data, i.e., $\mathcal{F}_{1:N} = \varnothing$. Thus the PM7 model and the training data are found to be mutually inconsistent. 

The inability to find a parameter vector feasible for all alkanes does not prove the non-existence of such a point. In fact, parameter values that are feasible for all alkanes could lead to potential biases in the prediction because methane and ethane are sufficiently different from the rest of the homologous series with respect to their atomic bonding and hence the energetics. Using parameter values feasible for methane and/or ethane in the prediction of larger alkane properties can result in a prediction interval with smaller uncertainty but in disagreement with experimental data, a problem similar to the bias-variance tradeoff.\cite{tibshirani2009elements}

Direct sampling the search region led to a fraction of samples being consistent with the experimental data for three alkanes (Table \ref{table:largestQOIpercentage}), 4250 samples were found from $\mathcal{F}_{6:8}$. Prediction of all alkanes by $\mathcal{F}_{6:8}$ is shown in Figure \ref{fig:predictionF68}. Feasible sets formed by consecutive alkanes, i.e., adjacent members of the homologous series, where only a few or no samples were found are shown in Table \ref{table:uniqueF}. The feasible set with the largest number of consecutive alkanes was $\mathcal{F}_{3:8}$. Prediction of all alkanes by $\mathcal{F}_{3:8}$ is shown in Figure \ref{fig:predictionF38}. No samples were found from $\mathcal{F}_{2:8}$ from direct sampling. Using a genetic algorithm via MATLAB's \textit{ga} function (and 750+ cpu hours), we were able to obtain a single parameter vector from $\mathcal{F}_{2:8}$ (Table \ref{table:geneticparameter}). 

\begin{figure}
	\centering
	\subfigure[]{
		\label{fig:lineplot_short}%
		\includegraphics[width=0.45\textwidth]{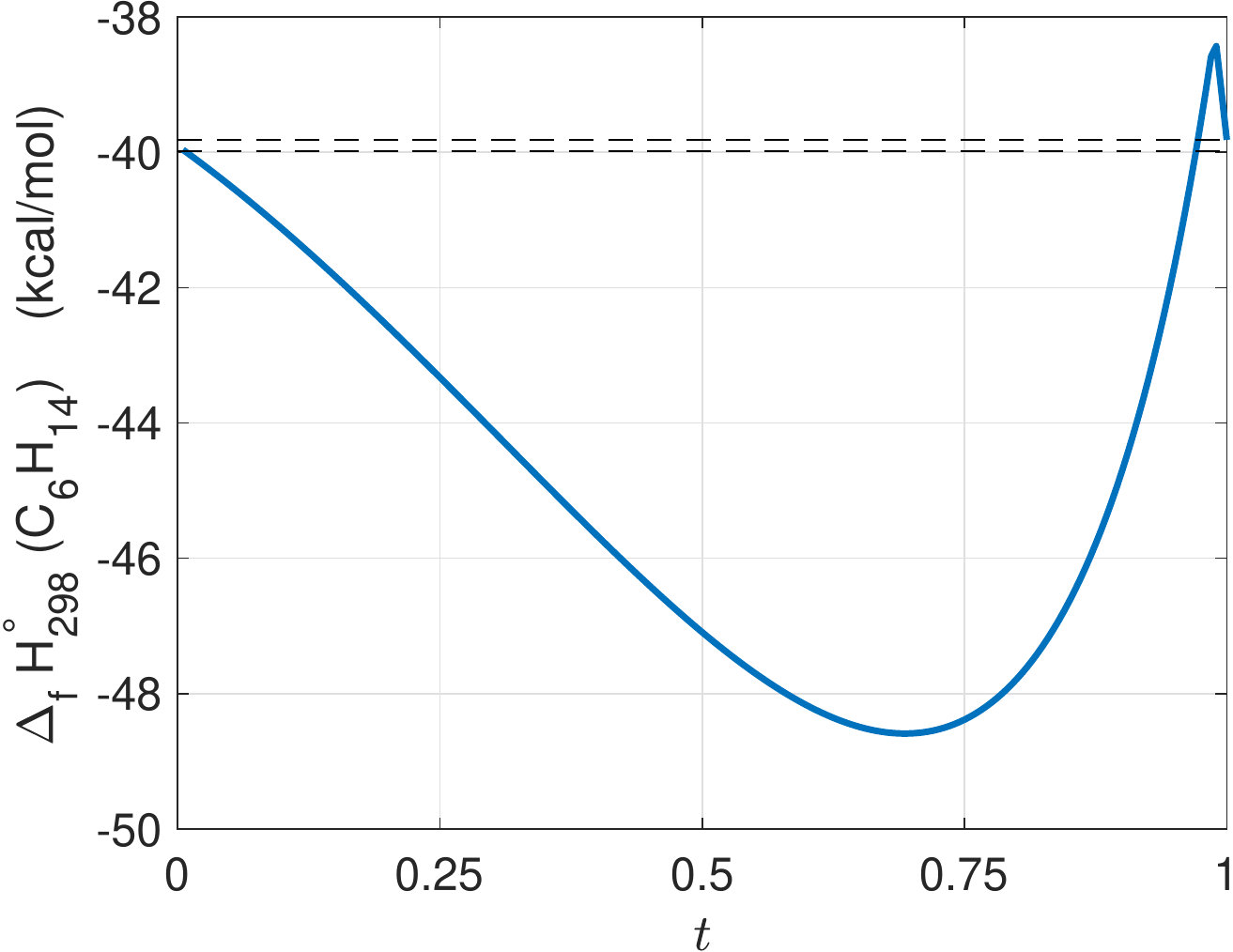}}%
	\qquad
	\subfigure[]{%
		\label{fig:lineplot_long}%
		\includegraphics[width=0.45\textwidth]{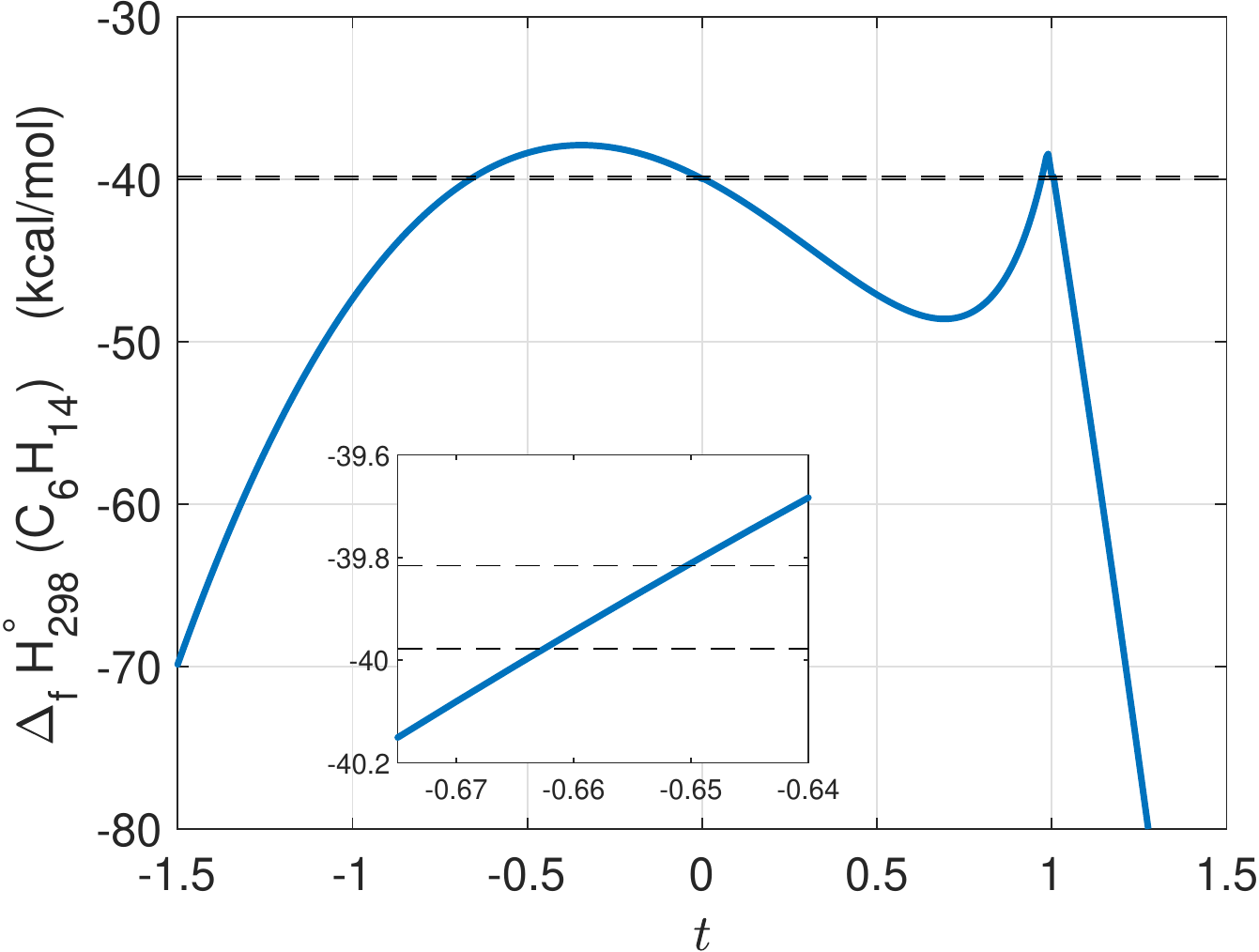}}%
	\caption{Blue curve is the heat of formation of \ce{C6H14} from PM7 calculations along the line segment defined as $x = (1 - t) x_1 + t x_2$, where $x_1$ and $x_2$ are two feasible points for \ce{C6H14}. The black dashed lines are the respective experimental bounds (see Table \ref{table:expData}). Panel A: line segment between two feasible points is not entirely feasible, thus the feasible set of \ce{C6H14} is non-convex. Panel B: extending the line segment beyond $x_1$, to the left, a new intersection with the experimental bounds is found at  $t = - 0.65$.}
	\label{fig:lineplot}
\end{figure}

To better understand the shape of a feasible set over the search region, consider two feasible points for \ce{C6H14}. The heat of formation of \ce{C6H14} was calculated along the line segment connecting the two feasible points. Figure \ref{fig:lineplot_short} shows that points along the line segment do not remain feasible, thus proving the feasible set is non-convex. Interestingly, if the line segment is extended further, beyond the initial sample region, another intersection with the experimental bounds occurs at $t = -0.65$, shown in Figure \ref{fig:lineplot_long}. This proves that there exist more feasible points for \ce{C6H14} which are outside of the sampled search region.

\subsection{Uncertainty of Predictions}

\begin{figure}[!htb]
	\centering
	\subfigure[]{
		\label{fig:c7predict}%
		\includegraphics[width=0.49\textwidth]{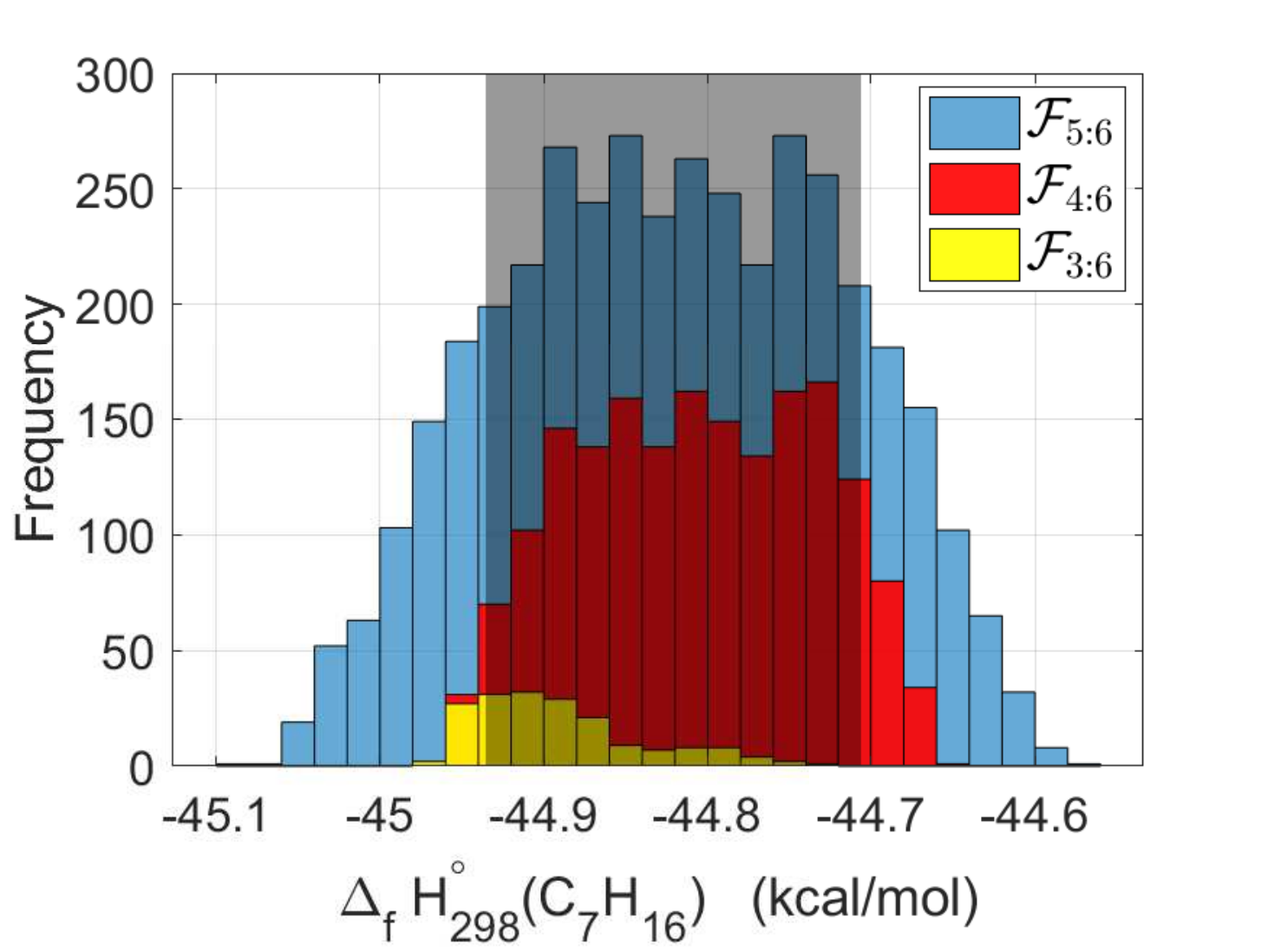}}%
	\hfill
	\subfigure[]{%
		\label{fig:c8predict}%
		\includegraphics[width=0.49\textwidth]{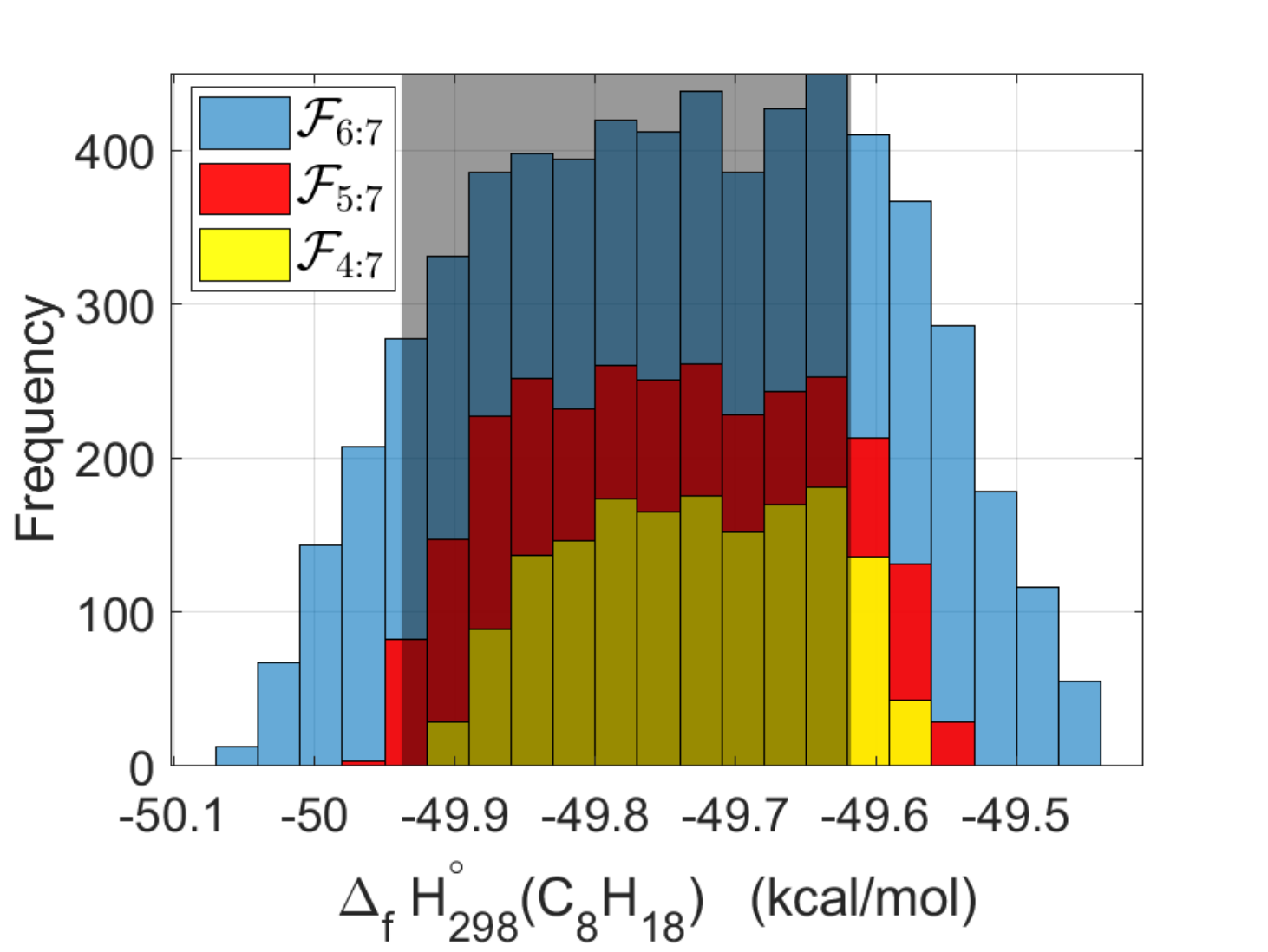}}%
	\qquad
	\subfigure[]{%
		\label{fig:c9predict}%
		\includegraphics[width=0.49\textwidth]{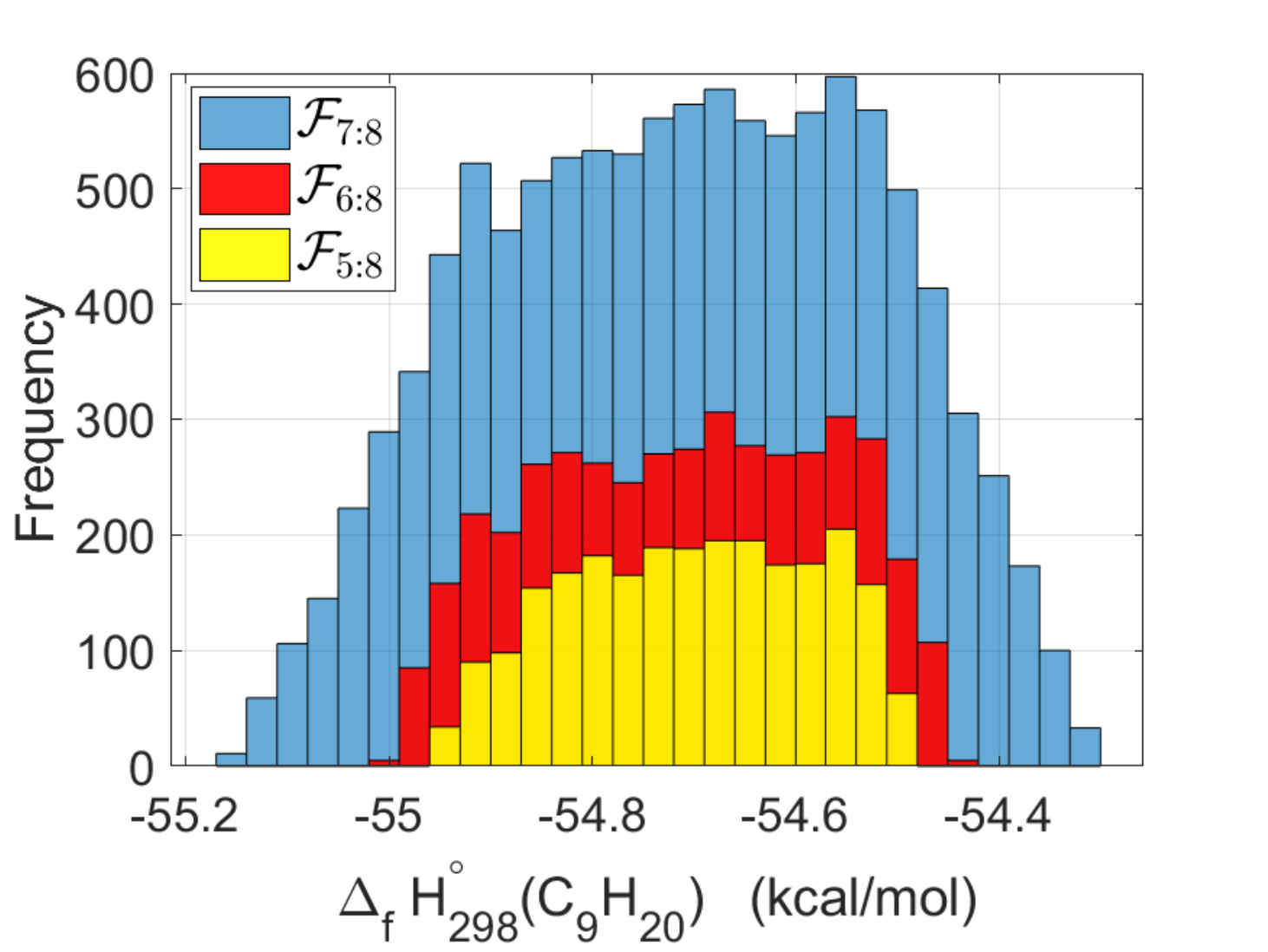}}%
	\caption{Heat of formation of a larger alkane, predicted from the feasible parameter set of smaller alkanes. The grey shaded region is the experimental interval for the respective alkane from Table \ref{table:expData}. Panel A: heat of formation of heptane predicted from samples of $\mathcal{F}_{5:6}$, $\mathcal{F}_{4:6}$  and $\mathcal{F}_{3:6}$. Panel B: heat of formation of octane predicted from samples of $\mathcal{F}_{6:7}$, $\mathcal{F}_{5:7}$, and $\mathcal{F}_{4:7}$. Panel C: heat of formation of nonane predicted from samples of $\mathcal{F}_{7:8}$, $\mathcal{F}_{6:8}$, and $\mathcal{F}_{5:8}$; this case is without experimental data.}
	\label{fig:prediction}
\end{figure}

In this section, we examine the uncertainty of predictions for larger alkanes using samples consistent with experimental data for smaller alkanes. Prediction of the heat of formation of heptane, octane, and nonane were investigated. In the case of heptane and octane, experimental data are available for comparison, but for nonane there are not. In each case, samples consistent with the experimental data for smaller alkanes were used for prediction of the larger alkane forming the histograms shown in Figure \ref{fig:prediction}. The width of the histograms constitute the prediction uncertainty. The predictions are only inner approximations as the samples were only generated from a portion of the parameter space. Thus, the predicted uncertainty can be no smaller than that which is presented below.

Figure \ref{fig:c7predict} depicts the prediction of the heat of formation of heptane. In blue are samples from $\mathcal{F}_{5:6}$, i.e, parameter vectors that were feasible for both \ce{C5H12} and \ce{C6H14}. The width of the distribution is 0.5 kcal/mol, which is larger than that of the experimental uncertainty of 0.2 kcal/mol (shown in grey), but still within the chemical accuracy (1 kcal/mol). To further reduce the prediction uncertainty, we can impose additional constraints by considering samples that are also feasible with smaller alkanes. Samples of $\mathcal{F}_{4:6}$ and $\mathcal{F}_{3:6}$, shown in red and yellow, respectively, are indeed capable of reducing the predicted uncertainty, and in the case of $\mathcal{F}_{3:6}$ producing a prediction 52.5\% smaller than that of $\mathcal{F}_{5:6}$. 

Following a similar analysis, we examined the prediction for octane, the largest alkane that we had experimental data for comparison. To predict the heat of formation of \ce{C8H18}, we considered feasible samples from the preceding two, three, and four smaller alkanes, i.e., samples from  $\mathcal{F}_{6:7}$, $\mathcal{F}_{5:7}$, and $\mathcal{F}_{4:7}$. The results are shown in Figure \ref{fig:c8predict}. The heat of formation of octane predicted by $\mathcal{F}_{6:7}$, shown in blue, yielded uncertainty of 0.62 kcal/mol, again within the chemical accuracy. Using samples which were feasible with the preceding three alkanes, $\mathcal{F}_{5:7}$, resulted in a prediction uncertainty of 0.42 kcal/mol, a reduction of 31\% compared to the prediction by $\mathcal{F}_{6:7}$. 

The computational results for the alkane without any experimental data for comparison, nonane, are shown in Figure \ref{fig:c9predict}. Using samples of $\mathcal{F}_{7:8}$, the prediction uncertainty for nonane is 0.85 kcal/mol. We were able to reduce the prediction uncertainty for nonane by 44\%, to 0.475 kcal/mol, by considering feasible samples from $\mathcal{F}_{5:8}$. Prediction of the heat of formation of nonane using $\mathcal{F}_{3:8}$, the feasible set with the largest number of consecutive alkanes found, is given in Figure \ref{fig:predictionc9All}. 

\begin{figure}[!htb]
	\centering\includegraphics[width=0.7\textwidth]{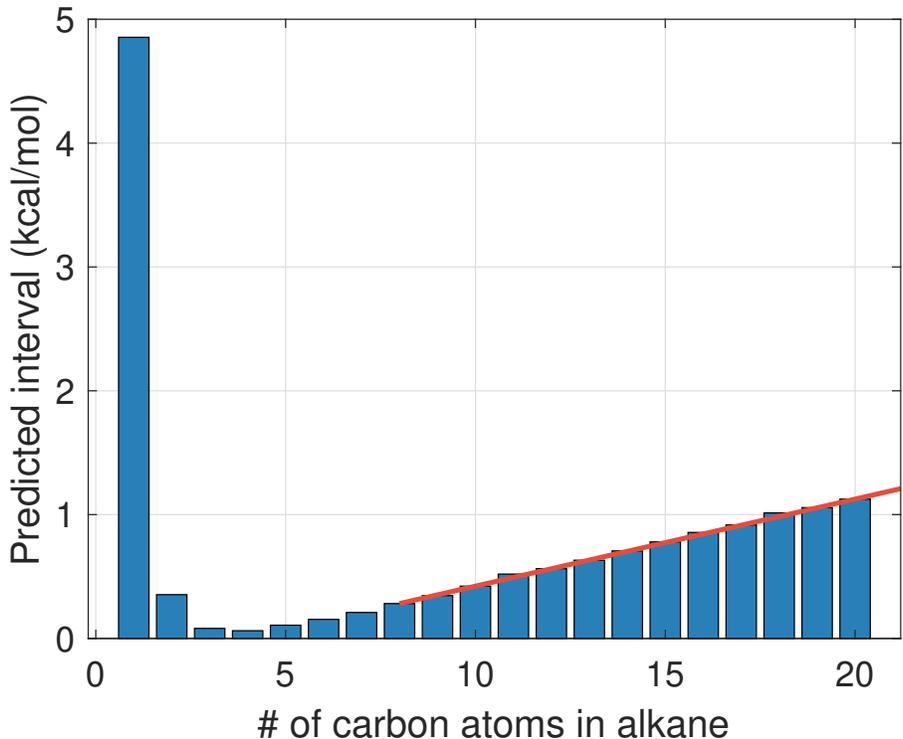}
	\caption{Predicted intervals for alkanes outside of the training data. Prediction was made using samples of $\mathcal{F}_{3:8}$. Interval length is defined as $\max_{x \in \mathcal{F}_{3:8}} M_p(x) - \min_{x \in \mathcal{F}_{3:8}} M_p(x)$, where $M_p(x)$ is the evaluation of PM7 of the $p$-th alkane with the parameter vector, $x$. The red line shown is a linear fit between n-\ce{C8H18} and n-\ce{C20H42}.}
	\label{fig:predictionIntervals}
\end{figure}

\begin{figure}[!htb]
	\centering\includegraphics[width=0.7\textwidth]{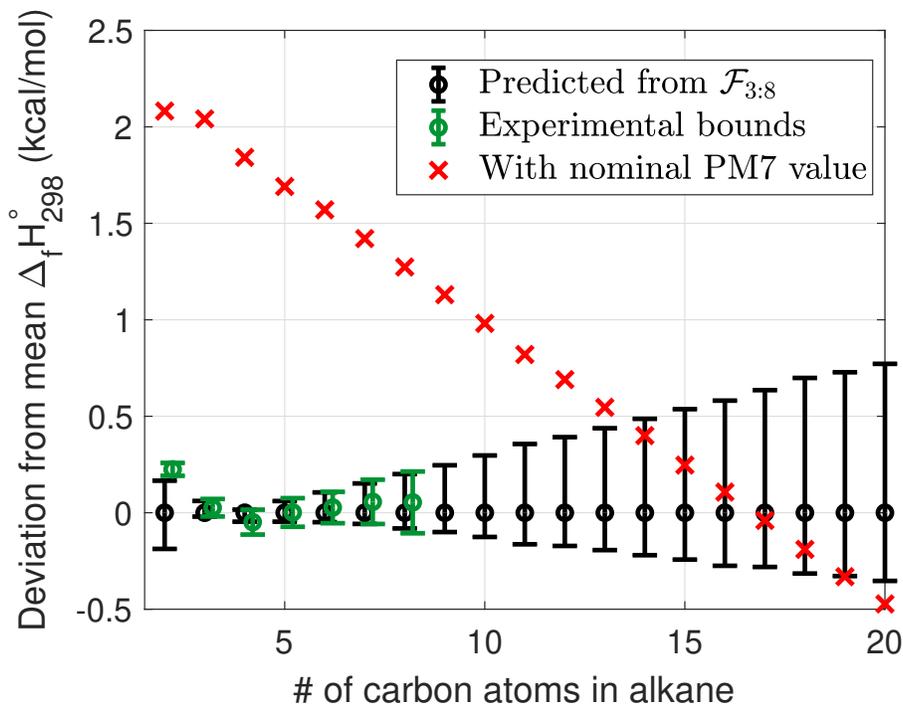}
	\caption{Uncertainty intervals of the heat of formation of ethane to icosane predicted by samples of $\mathcal{F}_{3:8}$. The black intervals are prediction by $\mathcal{F}_{3:8}$ and with the computed mean positioned at zero and marked with a black circle, the green intervals are the experimental bounds (Table \ref{table:expData}), and the red crosses are the heats of formation computed with the PM7 nominal parameter vector.}
	\label{fig:predictionErrorBars}
\end{figure}

The predictions for alkanes beyond nonane are shown in Figure \ref{fig:predictionIntervals}. The displayed uncertainty was predicted by using samples of $\mathcal{F}_{3:8}$. Two interesting observations can be made from these results. First, using samples from $\mathcal{F}_{3:8}$ produces a predicted uncertainty of nearly 5 kcal/mol for methane. This observation is consistent with the fact that methane is sufficiently different from the larger members of the homologous series. Parameter vectors feasible for larger alkanes are not capable of predicting the heat of formation for methane with chemical accuracy. Second, for larger alkanes prediction uncertainty grows linearly with the alkane size. The linearly increasing prediction uncertainty can also be seen in Figure \ref{fig:predictionErrorBars} where \ce{C2H6} through n-\ce{C20H42} are compared against the available experimental data and the nominal PM7 evaluation. Predictions made with samples of $\mathcal{F}_{3:8}$ overlap with the experimental bounds from \ce{C3H8} through \ce{C8H18} (by definition of the feasible set $\mathcal{F}_{3:8}$), but the nominal parameterization of PM7 deviates by nearly 2 kcal/mol. The PM7 results with the nominal parameter vector intersects with the predicted interval for alkanes \ce{C14H30} through \ce{C19H40}. 

\subsection{Classification of Feasible Points} 

As was established in Section \ref{section:fss} and Figure \ref{fig:lineplot}, the feasible set of PM7 is non-convex. For a convex feasible set it suffices to map out its boundary in order to determine if any combination of parameters is feasible or not. In the case of PM7, the feasible set is non-convex making an explicit mapping of the boundary of the domain computationally expensive; a viable alternative is to use the data collected via direct sampling and train a binary classifier that distinguishes ``feasible'' class of parameter values from the ``infeasible'' one. In this context, a feasible point could be associated with a specific feasibility label, such as $\mathcal{F}_{4:7}$, or any feasibility label. Given the unbalanced nature of the feasible and infeasible classes, i.e., over $5 \times 10^6$ samples with a infeasible label and only a few thousand samples belonging to any specific feasibility label, we choose the latter option and constructed the ``feasible'' class from all points that are feasible for at least one alkane in the training set. 

\begin{figure}[!htb]
	\centering
	\includegraphics[width=0.7\textwidth]{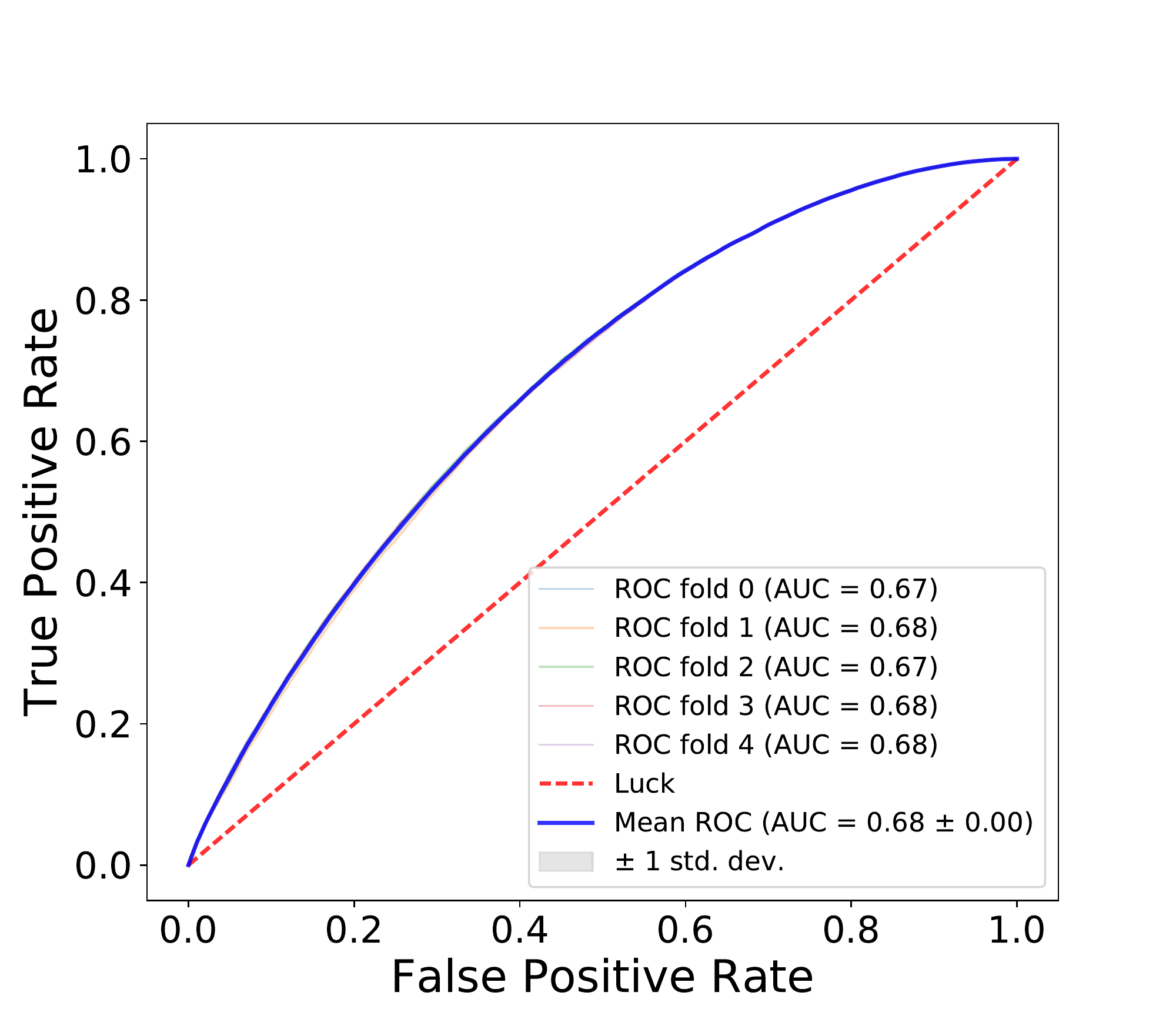}
	\caption{ROC and AUC for the binary classification (Random Forest model) of the feasibility of the points from the PM7 parameter space obtained via direct sampling. ``Feasible'' class includes points that are feasible for at least one training data point, ``Infeasible'' class includes points that are not feasible for any training data point. Both classes were down-sampled to $10^5$ samples. Classification performance was assessed by 5-fold cross-validation using $5 \times 10^3$ ensemble of random trees.}
	\label{fig:auc}
\end{figure}

Binary classification was performed using a Random Forest classifier, as implemented in Scikit-learn library.\cite{scikit-learn} A Random Forest classifier is capable of emulating the decision boundary for non-convex and even non-contiguous feasible domains. Both classes were down-sampled to $10^5$ to handle the class imbalance. Down-sampling was performed by using a randomly drawn $10^5$ samples from each class. Performance of the classifier is evaluated in a 5-fold cross-validation via construction of the Receiver Operating Curve (ROC) and computation of the Area Under Curve (AUC) (see Figure \ref{fig:auc}). 


With an AUC value of 0.68, we found that there is indeed a non-random structure in the feasible set recovered via direct sampling. Therefore, it is possible to train a classifier that fulfills the same function as the boundary to a feasible set.  Of course, for this assessment to be practical, a better performance metric of the classifier should be achieved. One route toward this goal is to collect a sample in the parameter space that conveys a better representation of the spatial extents of the non-convex feasible set found via sampling on a Latin hyper-cube. The inset of Figure \ref{fig:lineplot_long} depicts a region where a line segment passes through the experimental bounds. This region (and all other intersections) is a 27 dimensional space. Samples in these regions, near the experimental bounds, can help improve the characterization of the feasible set for classification.  

\section{Conclusion} 

Uncertainty quantification of semi-empirical quantum chemical model PM7 was performed using the B2BDC framework. The obtained results did not show evidence of model consistency even in the ``best case'' scenario, when we considered heats of formation of a small family of linear alkanes. This finding, however, requires a cautious interpretation. 

Lack of evidence of the model consistency with experimental data may imply that either the experimental data has some bias or the model has a deficiency. Given the quality of the experimental data, we tend to think the model inconsistency resides with the parameterization of PM7. To ultimately resolve this question, would require a significantly more engaged investigation of PM7 parameter space whose dimensionality exceeds 27 for organic molecules that include hetero-atoms. The practical lesson of the study is that we could not find a parameter vector feasible for all the training data having searched a finite volume and consumed a certain amount of the computational resources and time. The cost of establishing model consistency is, in itself, a diagnostic that can be used to compare various data-centric models and make an informed selection among them. 

Another practical lesson is the source of the encountered lack of consistency. The ``problematic'' QOIs found were the shortest alkanes, methane and ethane, whose chemical properties are known to be different from the rest of the homologous series. The value of PM7 comes from the low computational cost to evaluate larger molecules. In our study, the heats of formation of larger molecules were found to be consistency with the experimental data. Therefore, even if lack of PM7 consistency is rigorously proven at a later point, PM7 still can be treated as a consistent model for practical purposes and be used for prediction. The reported results demonstrate B2BDC's capability for a deeper analysis into PM7 by evaluating the limits of model consistency in a data-centric setting. 

One clear challenge in this study was the inability to accurately represent a significant volume of the parameter space with a surrogate model. The encountered difficulties motivated the development of a tool-set for uncertainty quantification of models with high-dimensional parameter spaces that can have non-convex and even non-contiguous feasible sets. Incorporation of machine learning techniques with the B2BDC framework was shown to be a viable strategy to efficiently tackle cases of this type.  Our basic attempt to learn the geometric structure of the identified feasible sets showed encouraging results. Possibility of the accurate binary classification of feasible/infeasible classes suggests a route to sampling strategies that are more efficient than brute-force direct sampling and could rely on a semi-supervised model to search for feasible points.

An equally important aspect of our study concerns uncertainty propagation from the training data that have sub-chemical accuracy to PM7 predictions. We observed that predictions remained within bounds of chemical accuracy for the molecules whose size is close to the size of the molecules in the training set. As expected, improvement of the quality of the feasible sets, i.e., attaining consistency with more alkanes, led to the reduction of the prediction uncertainty. Growth of the prediction error with the difference in the size between the test and training molecules agrees with the size-extensive nature of the selected QOI. This result reinforces intuition that there is a price to switching from interpolation to extrapolation regime. It motivates a large-scale uncertainty quantification study of semi-empirical quantum chemical models where both the size and chemical diversity of the dataset are significantly expanded. In order to facilitate such an effort, all parameter vectors and current feasibility labels are available from the authors upon request.


\section*{Acknowledgments}
	
Work at UCB was supported by the Department of Energy, National Nuclear Security Administration, under Award Number(s) $\text{DE-NA}0002375$. The views and opinions of authors expressed herein do not necessarily state or reflect those of the United States Government or any agency thereof.

\bibliography{uqPM7}	

\begin{thebibliography}{43}
\providecommand{\natexlab}[1]{#1}
\providecommand{\url}[1]{\texttt{#1}}
\expandafter\ifx\csname urlstyle\endcsname\relax
  \providecommand{\doi}[1]{doi: #1}\else
  \providecommand{\doi}{doi: \begingroup \urlstyle{rm}\Url}\fi

\bibitem[Miller et~al.(1990)Miller, Kee, and
  Westbrook]{doi:10.1146/annurev.pc.41.100190.002021}
James~A Miller, Robert~J Kee, and Charles~K Westbrook.
\newblock Chemical kinetics and combustion modeling.
\newblock \emph{Annual Review of Physical Chemistry}, 41\penalty0 (1):\penalty0
  345--387, 1990.

\bibitem[Battin-Leclerc et~al.(2011)Battin-Leclerc, Blurock, Bounaceur,
  Fournet, Glaude, Herbinet, Sirjean, and Warth]{C0CS00207K}
Frederique Battin-Leclerc, Edward Blurock, Roda Bounaceur, Rene Fournet,
  Pierre-Alexandre Glaude, Olivier Herbinet, Baptiste Sirjean, and V.~Warth.
\newblock Towards cleaner combustion engines through groundbreaking detailed
  chemical kinetic models.
\newblock \emph{Chem. Soc. Rev.}, 40:\penalty0 4762--4782, 2011.
\newblock \doi{10.1039/C0CS00207K}.
\newblock URL \url{http://dx.doi.org/10.1039/C0CS00207K}.

\bibitem[Frenklach(2007)]{frenklach2007transforming}
Michael Frenklach.
\newblock Transforming data into knowledge---process informatics for combustion
  chemistry.
\newblock \emph{Proceedings of the combustion Institute}, 31\penalty0
  (1):\penalty0 125--140, 2007.
\newblock \doi{10.1016/j.proci.2006.08.121}.

\bibitem[Lemkul et~al.(2016)Lemkul, Huang, Roux, and
  MacKerell]{doi:10.1021/acs.chemrev.5b00505}
Justin~A. Lemkul, Jing Huang, Benoît Roux, and Alexander~D. MacKerell.
\newblock An empirical polarizable force field based on the classical drude
  oscillator model: Development history and recent applications.
\newblock \emph{Chemical Reviews}, 116\penalty0 (9):\penalty0 4983--5013, 2016.
\newblock \doi{10.1021/acs.chemrev.5b00505}.
\newblock URL \url{https://doi.org/10.1021/acs.chemrev.5b00505}.
\newblock PMID: 26815602.

\bibitem[Warshel et~al.(2007)Warshel, Kato, and
  Pisliakov]{doi:10.1021/ct700127w}
Arieh Warshel, Mitsunori Kato, and Andrei~V. Pisliakov.
\newblock Polarizable force fields:  history, test cases, and prospects.
\newblock \emph{Journal of Chemical Theory and Computation}, 3\penalty0
  (6):\penalty0 2034--2045, 2007.
\newblock \doi{10.1021/ct700127w}.
\newblock URL \url{https://doi.org/10.1021/ct700127w}.
\newblock PMID: 26636199.

\bibitem[Christensen et~al.(2016)Christensen, Kubař, Cui, and
  Elstner]{doi:10.1021/acs.chemrev.5b00584}
Anders~S. Christensen, Tomáš Kubař, Qiang Cui, and Marcus Elstner.
\newblock Semiempirical quantum mechanical methods for noncovalent interactions
  for chemical and biochemical applications.
\newblock \emph{Chemical Reviews}, 116\penalty0 (9):\penalty0 5301--5337, 2016.
\newblock \doi{10.1021/acs.chemrev.5b00584}.
\newblock URL \url{https://doi.org/10.1021/acs.chemrev.5b00584}.
\newblock PMID: 27074247.

\bibitem[Yilmazer and Korth(2013)]{doi:10.1021/jp402719k}
Nusret~Duygu Yilmazer and Martin Korth.
\newblock Comparison of molecular mechanics, semi-empirical quantum mechanical,
  and density functional theory methods for scoring protein–ligand
  interactions.
\newblock \emph{The Journal of Physical Chemistry B}, 117\penalty0
  (27):\penalty0 8075--8084, 2013.
\newblock \doi{10.1021/jp402719k}.
\newblock URL \url{https://doi.org/10.1021/jp402719k}.
\newblock PMID: 23758433.

\bibitem[Walter(2013)]{doi:10.1002/wcms.1161}
Thiel Walter.
\newblock Semiempirical quantum–chemical methods.
\newblock \emph{Wiley Interdisciplinary Reviews: Computational Molecular
  Science}, 4\penalty0 (2):\penalty0 145--157, 2013.
\newblock \doi{10.1002/wcms.1161}.
\newblock URL \url{https://onlinelibrary.wiley.com/doi/abs/10.1002/wcms.1161}.

\bibitem[Burke(2012)]{doi:10.1063/1.4704546}
Kieron Burke.
\newblock Perspective on density functional theory.
\newblock \emph{The Journal of Chemical Physics}, 136\penalty0 (15):\penalty0
  150901, 2012.
\newblock \doi{10.1063/1.4704546}.
\newblock URL \url{https://doi.org/10.1063/1.4704546}.

\bibitem[Wiitala et~al.(2006)Wiitala, Hoye, and Cramer]{doi:10.1021/ct6001016}
Keith~W. Wiitala, Thomas~R. Hoye, and Christopher~J. Cramer.
\newblock Hybrid density functional methods empirically optimized for the
  computation of 13c and 1h chemical shifts in chloroform solution.
\newblock \emph{Journal of Chemical Theory and Computation}, 2\penalty0
  (4):\penalty0 1085--1092, 2006.
\newblock \doi{10.1021/ct6001016}.
\newblock URL \url{https://doi.org/10.1021/ct6001016}.
\newblock PMID: 26633067.

\bibitem[Karton et~al.(2008)Karton, Tarnopolsky, Lamère, Schatz, and
  Martin]{doi:10.1021/jp801805p}
Amir Karton, Alex Tarnopolsky, Jean-François Lamère, George~C. Schatz, and
  Jan M.~L. Martin.
\newblock Highly accurate first-principles benchmark data sets for the
  parametrization and validation of density functional and other approximate
  methods. derivation of a robust, generally applicable, double-hybrid
  functional for thermochemistry and thermochemical kinetics.
\newblock \emph{The Journal of Physical Chemistry A}, 112\penalty0
  (50):\penalty0 12868--12886, 2008.
\newblock \doi{10.1021/jp801805p}.
\newblock URL \url{https://doi.org/10.1021/jp801805p}.
\newblock PMID: 18714947.

\bibitem[Zhao et~al.(2006)Zhao, Schultz, and Truhlar]{doi:10.1021/ct0502763}
Yan Zhao, Nathan~E. Schultz, and Donald~G. Truhlar.
\newblock Design of density functionals by combining the method of constraint
  satisfaction with parametrization for thermochemistry, thermochemical
  kinetics, and noncovalent interactions.
\newblock \emph{Journal of Chemical Theory and Computation}, 2\penalty0
  (2):\penalty0 364--382, 2006.
\newblock \doi{10.1021/ct0502763}.
\newblock URL \url{https://doi.org/10.1021/ct0502763}.
\newblock PMID: 26626525.

\bibitem[Cui and Elstner(2014)]{C4CP00908H}
Qiang Cui and Marcus Elstner.
\newblock Density functional tight binding: values of semi-empirical methods in
  an ab initio era.
\newblock \emph{Phys. Chem. Chem. Phys.}, 16:\penalty0 14368--14377, 2014.
\newblock \doi{10.1039/C4CP00908H}.
\newblock URL \url{http://dx.doi.org/10.1039/C4CP00908H}.

\bibitem[Xue et~al.(2015)Xue, Dobbs, Bonvin, and
  Honavar]{doi:10.1016/j.febslet.2015.10.003}
Li~C Xue, Drena Dobbs, Alexandre~MJJ Bonvin, and Vasant Honavar.
\newblock Computational prediction of protein interfaces: A review of data
  driven methods.
\newblock \emph{FEBS letters}, 589\penalty0 (23):\penalty0 3516--3526, 2015.

\bibitem[Lusci et~al.(2013)Lusci, Pollastri, and Baldi]{doi:10.1021/ci400187y}
Alessandro Lusci, Gianluca Pollastri, and Pierre Baldi.
\newblock Deep architectures and deep learning in chemoinformatics: The
  prediction of aqueous solubility for drug-like molecules.
\newblock \emph{Journal of Chemical Information and Modeling}, 53\penalty0
  (7):\penalty0 1563--1575, 2013.
\newblock \doi{10.1021/ci400187y}.
\newblock URL \url{https://doi.org/10.1021/ci400187y}.
\newblock PMID: 23795551.

\bibitem[Sch{\"u}tt et~al.(2017)Sch{\"u}tt, Arbabzadah, Chmiela, M{\"u}ller,
  and Tkatchenko]{schutt2017quantum}
Kristof~T Sch{\"u}tt, Farhad Arbabzadah, Stefan Chmiela, Klaus~R M{\"u}ller,
  and Alexandre Tkatchenko.
\newblock Quantum-chemical insights from deep tensor neural networks.
\newblock \emph{Nature communications}, 8:\penalty0 13890, 2017.

\bibitem[Hegde and Bowen(2017)]{hegde2017machine}
Ganesh Hegde and R~Chris Bowen.
\newblock Machine-learned approximations to density functional theory
  hamiltonians.
\newblock \emph{Scientific Reports}, 7:\penalty0 42669, 2017.

\bibitem[Goh et~al.(2017)Goh, Hodas, and Vishnu]{doi:10.1002/jcc.24764}
Garrett~B Goh, Nathan~O Hodas, and Abhinav Vishnu.
\newblock Deep learning for computational chemistry.
\newblock \emph{Journal of Computational Chemistry}, 38\penalty0 (16):\penalty0
  1291--1307, 2017.
\newblock \doi{10.1002/jcc.24764}.
\newblock URL \url{https://onlinelibrary.wiley.com/doi/abs/10.1002/jcc.24764}.

\bibitem[McGibbon et~al.(2017)McGibbon, Taube, Donchev, Siva, Hernández,
  Hargus, Law, Klepeis, and Shaw]{doi:10.1063/1.4986081}
Robert~T. McGibbon, Andrew~G. Taube, Alexander~G. Donchev, Karthik Siva, Felipe
  Hernández, Cory Hargus, Ka-Hei Law, John~L. Klepeis, and David~E. Shaw.
\newblock Improving the accuracy of møller-plesset perturbation theory with
  neural networks.
\newblock \emph{The Journal of Chemical Physics}, 147\penalty0 (16):\penalty0
  161725, 2017.
\newblock \doi{10.1063/1.4986081}.
\newblock URL \url{https://doi.org/10.1063/1.4986081}.

\bibitem[Medvedev et~al.(2017)Medvedev, Bushmarinov, Sun, Perdew, and
  Lyssenko]{medvedev2017density}
Michael~G Medvedev, Ivan~S Bushmarinov, Jianwei Sun, John~P Perdew, and
  Konstantin~A Lyssenko.
\newblock Density functional theory is straying from the path toward the exact
  functional.
\newblock \emph{Science}, 355\penalty0 (6320):\penalty0 49--52, 2017.

\bibitem[Cherkasov et~al.(2014)Cherkasov, Muratov, Fourches, Varnek, Baskin,
  Cronin, Dearden, Gramatica, Martin, Todeschini, Consonni, Kuz’min, Cramer,
  Benigni, Yang, Rathman, Terfloth, Gasteiger, Richard, and
  Tropsha]{doi:10.1021/jm4004285}
Artem Cherkasov, Eugene~N. Muratov, Denis Fourches, Alexandre Varnek, Igor~I.
  Baskin, Mark Cronin, John Dearden, Paola Gramatica, Yvonne~C. Martin, Roberto
  Todeschini, Viviana Consonni, Victor~E. Kuz’min, Richard Cramer, Romualdo
  Benigni, Chihae Yang, James Rathman, Lothar Terfloth, Johann Gasteiger, Ann
  Richard, and Alexander Tropsha.
\newblock Qsar modeling: Where have you been? where are you going to?
\newblock \emph{Journal of Medicinal Chemistry}, 57\penalty0 (12):\penalty0
  4977--5010, 2014.
\newblock \doi{10.1021/jm4004285}.
\newblock URL \url{https://doi.org/10.1021/jm4004285}.
\newblock PMID: 24351051.

\bibitem[Mansouri et~al.(2016)Mansouri, Grulke, Richard, Judson, and
  Williams]{doi:10.1080/1062936X.2016.1253611}
K.~Mansouri, C.~M. Grulke, A.~M. Richard, R.~S. Judson, and A.~J. Williams.
\newblock An automated curation procedure for addressing chemical errors and
  inconsistencies in public datasets used in qsar modelling.
\newblock \emph{SAR and QSAR in Environmental Research}, 27\penalty0
  (11):\penalty0 911--937, 2016.
\newblock \doi{10.1080/1062936X.2016.1253611}.
\newblock URL \url{https://doi.org/10.1080/1062936X.2016.1253611}.
\newblock PMID: 27885862.

\bibitem[Peverati and Truhlar(2014)]{peverati2014quest}
Roberto Peverati and Donald~G Truhlar.
\newblock Quest for a universal density functional: the accuracy of density
  functionals across a broad spectrum of databases in chemistry and physics.
\newblock \emph{Phil. Trans. R. Soc. A}, 372\penalty0 (2011):\penalty0
  20120476, 2014.

\bibitem[Simm et~al.(2017)Simm, Proppe, and Reiher]{Simm:2017:0009-4293:202}
Gregor~N. Simm, Jonny Proppe, and Markus Reiher.
\newblock Error assessment of computational models in chemistry.
\newblock \emph{CHIMIA International Journal for Chemistry}, 71\penalty0
  (4):\penalty0 202--208, 2017.
\newblock ISSN 0009-4293.
\newblock \doi{doi:10.2533/chimia.2017.202}.
\newblock URL
  \url{https://www.ingentaconnect.com/content/scs/chimia/2017/00000071/00000004/art00011}.

\bibitem[Russi et~al.(2010)Russi, Packard, and Frenklach]{russi2010uncertainty}
Trent Russi, Andy Packard, and Michael Frenklach.
\newblock Uncertainty quantification: Making predictions of complex reaction
  systems reliable.
\newblock \emph{Chemical Physics Letters}, 499\penalty0 (1-3):\penalty0 1--8,
  2010.

\bibitem[Faver et~al.(2012)Faver, Yang, and Merz~Jr]{faver2012effects}
John~C Faver, Wei Yang, and Kenneth~M Merz~Jr.
\newblock The effects of computational modeling errors on the estimation of
  statistical mechanical variables.
\newblock \emph{Journal of Chemical Theory and Computation}, 8\penalty0
  (10):\penalty0 3769--3776, 2012.

\bibitem[Yang et~al.(2018)Yang, Lei, Gao, Thomas, Mobley, and
  Baker]{yang2018atomic}
Xiu Yang, Huan Lei, Peiyuan Gao, Dennis~G. Thomas, David~L. Mobley, and
  Nathan~A. Baker.
\newblock Atomic radius and charge parameter uncertainty in biomolecular
  solvation energy calculations.
\newblock \emph{Journal of Chemical Theory and Computation}, 14\penalty0
  (2):\penalty0 759--767, 2018.
\newblock \doi{10.1021/acs.jctc.7b00905}.
\newblock URL \url{https://doi.org/10.1021/acs.jctc.7b00905}.
\newblock PMID: 29293342.

\bibitem[Simm and Reiher(2018)]{simm2018error}
Gregor~N Simm and Markus Reiher.
\newblock Error-controlled exploration of chemical reaction networks with
  gaussian processes.
\newblock \emph{arXiv preprint arXiv:1805.09886}, 2018.

\bibitem[Edwards et~al.(2014)Edwards, Zubarev, Packard, Lester, and
  Frenklach]{PhysRevLett.112.253003}
David~E. Edwards, Dmitry~Yu. Zubarev, Andrew Packard, William~A. Lester, and
  Michael Frenklach.
\newblock Interval prediction of molecular properties in parametrized quantum
  chemistry.
\newblock \emph{Phys. Rev. Lett.}, 112:\penalty0 253003, Jun 2014.
\newblock \doi{10.1103/PhysRevLett.112.253003}.
\newblock URL \url{https://link.aps.org/doi/10.1103/PhysRevLett.112.253003}.

\bibitem[Frenklach et~al.(2002)Frenklach, Packard, and
  Seiler]{frenklach2002prediction}
Michael Frenklach, Andrew Packard, and Pete Seiler.
\newblock Prediction uncertainty from models and data.
\newblock In \emph{Proceedings of the American Control Conference}, volume~5,
  pages 4135--4140. IEEE, 2002.

\bibitem[Seiler et~al.(2006)Seiler, Frenklach, Packard, and
  Feeley]{seiler2006numerical}
Pete Seiler, Michael Frenklach, Andrew Packard, and Ryan Feeley.
\newblock Numerical approaches for collaborative data processing.
\newblock \emph{Optimization and Engineering}, 7\penalty0 (4):\penalty0
  459--478, 2006.

\bibitem[Stewart(2013)]{stewart2013optimization}
James~JP Stewart.
\newblock Optimization of parameters for semiempirical methods vi: more
  modifications to the nddo approximations and re-optimization of parameters.
\newblock \emph{Journal of Molecular Modeling}, 19\penalty0 (1):\penalty0
  1--32, 2013.

\bibitem[Dral et~al.(2016{\natexlab{a}})Dral, Wu, Spörkel, Koslowski, Weber,
  Steiger, Scholten, and Thiel]{dral2016semiempirical1}
Pavlo~O Dral, Xin Wu, Lasse Spörkel, Axel Koslowski, Wolfgang Weber, Rainer
  Steiger, Mirjam Scholten, and Walter Thiel.
\newblock Semiempirical quantum-chemical orthogonalization-corrected methods:
  Theory, implementation, and parameters.
\newblock \emph{Journal of Chemical Theory and Computation}, 12\penalty0
  (3):\penalty0 1082--1096, 2016{\natexlab{a}}.

\bibitem[Dral et~al.(2016{\natexlab{b}})Dral, Wu, Spörkel, Koslowski, and
  Thiel]{dral2016semiempirical}
Pavlo~O Dral, Xin Wu, Lasse Spörkel, Axel Koslowski, and Walter Thiel.
\newblock Semiempirical quantum-chemical orthogonalization-corrected methods:
  Benchmarks for ground-state properties.
\newblock \emph{Journal of Chemical Theory and Computation}, 12\penalty0
  (3):\penalty0 1097--1120, 2016{\natexlab{b}}.

\bibitem[Feeley et~al.(2004)Feeley, Seiler, Packard, and
  Frenklach]{feeley2004consistency}
Ryan Feeley, Pete Seiler, Andrew Packard, and Michael Frenklach.
\newblock Consistency of a reaction dataset.
\newblock \emph{The Journal of Physical Chemistry A}, 108\penalty0
  (44):\penalty0 9573--9583, 2004.

\bibitem[Frenklach et~al.(2016)Frenklach, Packard, Garcia-Donato, Paulo, and
  Sacks]{frenklach2016comparison}
Michael Frenklach, Andrew Packard, Gonzalo Garcia-Donato, Rui Paulo, and Jerome
  Sacks.
\newblock Comparison of statistical and deterministic frameworks of uncertainty
  quantification.
\newblock \emph{SIAM/ASA Journal on Uncertainty Quantification}, 4\penalty0
  (1):\penalty0 875--901, 2016.

\bibitem[Box and Draper(1987)]{box1987empirical}
George~EP Box and Norman~R Draper.
\newblock \emph{Empirical model-building and response surfaces.}
\newblock John Wiley \& Sons, 1987.

\bibitem[Ruscic(2017)]{ruscic2017active}
B~Ruscic.
\newblock Active thermochemical tables (atct) values based on ver. 1.122 of the
  thermochemical network, 2016.
\newblock \emph{available at \url{ATcT.anl.gov}}, 2017.

\bibitem[MATLAB(2002)]{matlab}
MATLAB.
\newblock Statistics and machine learning toolbox, parallel computing toolbox,
  and optimization toolbox release 2017b, the mathworks inc.
\newblock \emph{Natick, MA}, 2002.

\bibitem[Stewart(2012)]{stewart2012colorado}
James~JP Stewart.
\newblock Mopac2016. stewart computational chemistry, colorado springs, co,
  usa.
\newblock \emph{available at \url{http://openmopac.net}}, 2012.

\bibitem[Benson et~al.(1969)Benson, Cruickshank, Golden, Haugen, O'neal,
  Rodgers, Shaw, and Walsh]{benson1969additivity}
Sidney~W Benson, FR~Cruickshank, DM~Golden, Gilbert~R Haugen, HE~O'neal,
  AS~Rodgers, Robert Shaw, and R~Walsh.
\newblock Additivity rules for the estimation of thermochemical properties.
\newblock \emph{Chemical Reviews}, 69\penalty0 (3):\penalty0 279--324, 1969.

\bibitem[Tibshirani et~al.(2009)Tibshirani, Friedman, and
  Hastie]{tibshirani2009elements}
R~Tibshirani, J~Friedman, and T~Hastie.
\newblock The elements of statistical learning, vol. 2, 2009.

\bibitem[Pedregosa et~al.(2011)Pedregosa, Varoquaux, Gramfort, Michel, Thirion,
  Grisel, Blondel, Prettenhofer, Weiss, Dubourg, Vanderplas, Passos,
  Cournapeau, Brucher, Perrot, and Duchesnay]{scikit-learn}
F.~Pedregosa, G.~Varoquaux, A.~Gramfort, V.~Michel, B.~Thirion, O.~Grisel,
  M.~Blondel, P.~Prettenhofer, R.~Weiss, V.~Dubourg, J.~Vanderplas, A.~Passos,
  D.~Cournapeau, M.~Brucher, M.~Perrot, and E.~Duchesnay.
\newblock Scikit-learn: Machine learning in {P}ython.
\newblock \emph{Journal of Machine Learning Research}, 12:\penalty0 2825--2830,
  2011.

\end{thebibliography}
\bibliographystyle{unsrtnat}
\end{document}